\documentclass[twocolumn,english,aps,superscriptaddress]{revtex4-1}
\pdfoutput=1 
\usepackage[T1]{fontenc}
\usepackage[utf8]{inputenc}
\usepackage{amsmath}
\usepackage{amssymb}
\usepackage{graphicx}
\usepackage{babel}
\begin{document}

\title{Inhibition as a determinant of activity and criticality in dynamical networks}

\author{Joao Pinheiro Neto}
\email{joaoxp@gmail.com}
\affiliation{Institute of Physics\emph{ ``Gleb Wataghin'', }University of Campinas, 13083-859, Campinas, Brazil}
\affiliation{Max Planck Institute for Dynamics and Self-Organization, 37077, Göttingen, Germany}
\author{Marcus A. M. de Aguiar}
\affiliation{Institute of Physics\emph{ ``Gleb Wataghin'', }University of Campinas, 13083-859, Campinas, Brazil}
\author{Jos\'e A. Brum}
\affiliation{Institute of Physics\emph{ ``Gleb Wataghin'', }University of Campinas, 13083-859, Campinas, Brazil}
\author{Stefan Bornholdt}
\affiliation{Institute for Theoretical Physics, University of Bremen, 28359, Bremen, Germany}

\begin{abstract}
A certain degree of inhibition is a common trait of dynamical 
networks in nature, ranging from neuronal and biochemical 
networks, to social and technological networks. 
We study here the role of inhibition in a representative dynamical 
network model, characterizing the dynamics of random threshold networks 
with both excitatory and inhibitory links. 
Varying the fraction of excitatory links has a strong effect on
the network's population activity and its sensitivity to perturbation. 
The average degree $K$, known to have a strong effect on the dynamics 
when small, loses its influence on the dynamics as its value increases.
Instead, the strength of inhibition is a determinant of 
dynamics and sensitivity here, allowing for criticality only in a specific 
corridor of inhibition. This criticality corridor requires that excitation dominates, 
while the balance region corresponds to maximum sensitivity to perturbation.
We develop mean-field approximations of the population activity and
sensitivity and find that the network dynamics is independent of degree 
distribution for high $K$. In a minimal model of an adaptive threshold network 
we demonstrate how the dynamics remains robust against changes in the topology.
This adaptive model can be extended in order to generate networks with 
a controllable activity distribution and specific topologies.
\end{abstract}
\maketitle

\section{Introduction}

Inhibition is a frequent factor in many real-world networks as nodes 
often have an excitatory, as well as an inhibitory effect on their neighbors. 
Actors in a social network may display affection or animosity towards each other, 
while genes may suppress the expression of others in a gene regulatory network. 
Notably, an estimate $20\%$ of the neurons in the brain are inhibitory, and have
a key role in regulating neuronal activity
\cite{Buzsaki2007, Markram2004a, Isaacson2011, Toyoizumi2013}.
In model systems, dynamics with inhibition can generate more complex
dynamics than their excitation-only counterparts \cite{Larremore2014}.

Another common feature in real-world networks, also present in the brain, is a topology that evolves with time. In adaptive
networks, feedback loops between network dynamics and topological
evolution can drive the network towards different dynamical states. 
These can be critical points of a phase transition
\cite{Bak1987, Bornholdt2000, Rybarsch2014}
or dynamical states not accessible to networks with standard (e.g.
lattice or random) topologies 
\cite{Gross2009, Gross2008a, Gross2006, Aoki2012}.
To prevent changes in the topology from resulting in catastrophic failure
of network function, it is important that the network dynamics be
robust against such changes. 
The observation of avalanches of activity in cortical tissue, pointing towards a dynamical state near a phase transition \cite{Beggs2003, Friedman2012, Plenz2014, Priesemann2014, Wilting2016}, has also sparked renewed interest in models with phase transitions and mechanisms capable of controlling network evolution.

Random threshold networks (RTNs) have been used to model a vast array
of phenomena \cite{Aldana2011,Andrecut2009,Rohlf2002,Rohlf2007,Szejka2008,Wang2013},
from neural networks \cite{McCulloch1943} to gene regulatory networks \cite{Davidich2008a,Li2004}.
While RTNs have been extensively studied, in most models the links
$w_{ij}$ between nodes are either always excitatory ($w_{ij}>0$)
or $w_{ij}=\pm1$ with equal probability. 
In the following we generalize the RTN model to a varying balance between excitatory and inhibitory links
in order to obtain insights into how inhibition may affect real-world networks. 
We use both network simulations (Sec. II) and analytic methods (Sec. III), and focus on the conditions for criticality and how we can control the network's activity and dynamical sensitivity. 
We also study its robustness to topological evolution, and how, by evolving the topology, a higher degree of control of the dynamics is obtained.

\begin{figure*}[!t]
\begin{centering}
\includegraphics[width=17.2cm]{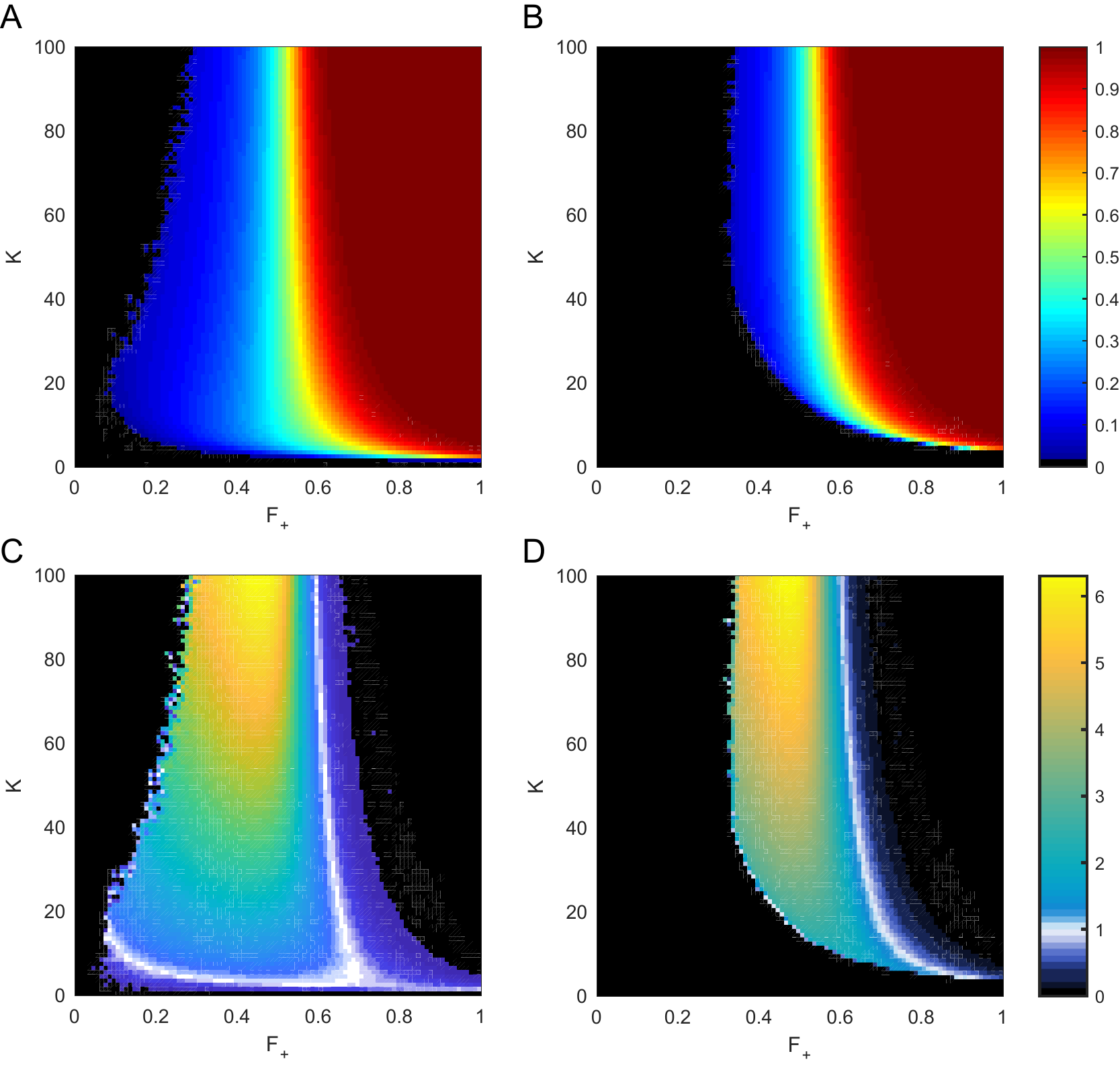}
\par\end{centering}

\centering{}\caption{\label{fig:ParamSpace}\textbf{Parameter space of the dynamics as
a function of $K$ and $F_{+}$.}
\textbf{A.} Stable activity $A_{\infty}$ for a threshold of $h=0$. 
\textbf{B.} Same as (A) for $h=1$. 
\textbf{C.} Sensitivity $\lambda$ for $h=0$. The white region corresponds to
networks near criticality ($\lambda=1$). 
\textbf{D.} Same as (C) for $h=1$. 
Balance ($F_+\approx0.5$) results in maximum sensitivity, while criticality requires more excitation ($F_+>0.5$) for high $K$.
Other parameters are $n=10^{3}$ and $A(0)=0.9$, and
each point is the average of $10^{2}$ simulations.}
\end{figure*}

\section{Dynamics of RTNs with inhibition}

\subsection{Model definition}

Consider a directed network with $n$ randomly connected nodes and
average degree $K$. A weight $w_{ij}=\pm1$ is assigned to each existing
link from node $j$ to node $i$. The dynamical state of each node
is represented by a Boolean variable. The state $\sigma_{i}(t+1)$
of node $i$ at time $t+1$ is given by
\begin{equation}
\sigma_{i}(t+1)=\begin{cases}
1 & \mbox{, if \ensuremath{\sum_{j=1}^{n}w_{ij}\sigma_{j}(t)>h}}\\
0 & \mbox{, if \ensuremath{\sum_{j=1}^{n}w_{ij}\sigma_{j}(t)\leq h}}
\end{cases}\label{eq:def-rtn}
\end{equation}
where $h$ is a threshold parameter. The network collective state
is measured by the fraction of active nodes,

\begin{equation}
A(t)=\frac{1}{n}\sum_{i}\sigma_{i}(t)\label{eq:def A(t)}
\end{equation}
The dynamics is initiated by randomly choosing a fraction of the nodes to
be activated at $t=0$, $A(0)$. We need a set of parameters to characterize
the relevant topological properties of the network. These are the
average degree $K$ of the network, and the fraction of positive links
$F_{+}$. The link weights $w_{ij}=\pm1$ are distributed randomly
in the network, with the constraint that the total fraction of positive
links is $F_{+}$. Unless otherwise stated, the underlying topology
is that of a directed Erdős–Rényi (ER) network. As
we will see, properties such as degree distribution are not critical
in the high connectivity regime.

Several variants for the RTNs are possible. Some works in the literature
add a self-regulating state \cite{Szejka2008,Aldana2011} to Eq.\ \ref{eq:def-rtn}
given by $\sigma_{i}\left(t+1\right)=\sigma_{i}\left(t\right)$ if
$\sum_{j=1}^{n}w_{ij}\sigma_{j}\left(t\right)=h$. This new condition
can drastically change the dynamics by adding long-term temporal correlations
between the node states. Note that for non-integer $h$ (and $w_{ij}=\pm1$)
this state is not possible, and the model is equivalent to ours. 
Our RTN definition with threshold $h=1$ is equivalent to the 
``biological'' RTN variant as defined in \cite{Rybarsch2012},
particularly suited for simulating genetic networks. 
Another possibility is to update the node states asynchronously \cite{Wang2013}
or use other forms of asynchronicity. 
In order to focus on the effect of $F_{+}$ in ensembles of random networks, 
we will use the simpler and most common case where all nodes are updated 
in parallel according to Eq.\ \ref{eq:def-rtn}.

\subsection{Parameter space}

Since the dynamics is deterministic it must eventually arrive at an
attractor. This can be a limit circle of $A\left(t\right)$, with
$A(t+\tau)=A(t)$ for some $\tau$, or a fixed point $A_{\infty}$,
with $A_{\infty}=A\left(t\rightarrow\infty\right)$. The value $A_{\infty}=0$
is always a fixed point of the dynamics, since $\sigma_{i}=0$ for
all $i$ is an absorbing state. Szejka et al. \cite{Szejka2008} showed
that for the specific case of $F_{+}=0.5$ another stable fixed point
$A_{\infty}>0$ can emerge depending on the relationship between $K$
and $h$. 

We ran numerical simulations in order to obtain $A_{\infty}$ for
the entire range $0<F_{+}<1$. In Fig.\ \ref{fig:ParamSpace}A,B we
plot the fixed point $A_{\infty}$ in the parameter space $A_{\infty}=A_{\infty}\left(F_{+},K\right)$
for fixed thresholds $h\in\{0,1\}$. We observe a much richer dynamics
than in the $F_{+}=0.5$ and $F_{+}=1$ cases. Intuitively, $A_{\infty}$
increases with $F_{+}$. However, the interesting aspect is that $A_{\infty}$
can have a value on a large range, with the crucial parameter being
$F_{+}$. The dependence on $K\in[10,100]$ is weaker, with the
wider range of $F_{+}$ with non-frozen dynamics (i.e., $A_{\infty}\neq0\mbox{ and }1$)
happening for lower $K$. A critical degree $K_{C}$ is necessary
for $A_{\infty}>0$, however.
As $K$ increases, the non-frozen dynamics becomes more centered around
$F_{+}=0.5$, and in the $K\rightarrow\infty$ limit only happens
for $A_{\infty}\left(0.5,K\rightarrow\infty\right)=0.5$. 
The convergence is slow, however.
As the threshold $h$ increases, the region in the $\left(F_{+},K\right)$ parameter
space with non-frozen dynamics decreases. It is important to note
that the transition from $A_{\infty}=0$ to $A_{\infty}>0$ is
not a continuous transition. Very low activity network states can
in principle happen, but they have a high probability of going to
the absorbing state $A(t)=0$ (see Appendix A). Thus, in practice
a finite network has a minimum stable activity $A_{\infty}^{C}$ that
it can sustain, which increases with larger $h$. In Sec. IIC we explore
$A_{\infty}^{C}$ and other observables that define the dynamic range
of RTNs.

Another important observable of RTNs is the network sensitivity $\lambda$.
The sensitivity is defined as the number of perturbed states at time
$t+1$ after changing the state of one node at time $t$, averaged
over an ensemble of network states \cite{Szejka2008,Aldana2011}. For
memoryless network dynamics this is enough to
determine how a perturbation will spread over long timescales. If
$\lambda>1$, the network is chaotic and a perturbation will spread
and take over the network. If $\lambda<1$, the network is ordered
and a perturbation will quickly die out. Finally, if $\lambda=1$,
the network is critical and a perturbation will take a long time to
disappear without dominating the dynamics. In Figs.\ \ref{fig:ParamSpace}C,D we show the parameter
space of $\lambda$ as a function of $F_{+}$ and $K$ for $h\in\{0,1\}$.
In general, $\lambda$ increases with $K$ and is highest for networks
with $F_{+}\approx0.5$. 

Of particular interest are the regions of criticality where $\lambda\approx1$.
In Fig.\ \ref{fig:ParamSpace}C the near-horizontal white line corresponds 
to the well known critical point of discrete dynamical networks for 
a small critical value of $K$ that only slightly depends on the,
otherwise nearly arbitrary, degree of inhibition. 
A main observation of our paper is the second, almost vertical, 
white line, indicating a second region of criticality for almost 
arbitrary and also high values of $K$, given that the fraction of 
positive links $F_{+}$ lies within a narrow, well defined region. 
Here, the critical state requires $<F_{+}>0.5$, asymptotically approaching the balanced state ($F_+ = 0.5$) in the $K\rightarrow\infty$ limit.
Nevertheless, well-connected finite-sized networks require more excitation than inhibition to     attain criticality.
Note also that, while criticality requires $<F_{+}>0.5$, $<F_{+}\approx0.5$ results in maximum sensitivity to perturbations (supercriticality).

Comparing the activity and sensitivity plots of Fig\ \ref{fig:ParamSpace}, 
we observe that the critical case is in general
possible for a narrow region in connectivity $K$ with relatively low activity,
and a broad range of $K$ at a higher activity level (see also Fig.\ \ref{fig:sensitivity}B).
This latter case, interestingly, combines conditions which are relevant for 
information processing networks: an intermediate activity level and 
a critical level of sensitivity. That this is closely correlated to a 
narrow range of inhibition in our RTN model may have implications for 
similar relationships in natural networks, as in the case of inhibition 
in the brain.

\subsection{Dynamic range of RTNs}

\begin{figure}
\begin{centering}
\includegraphics[width=8.6cm]{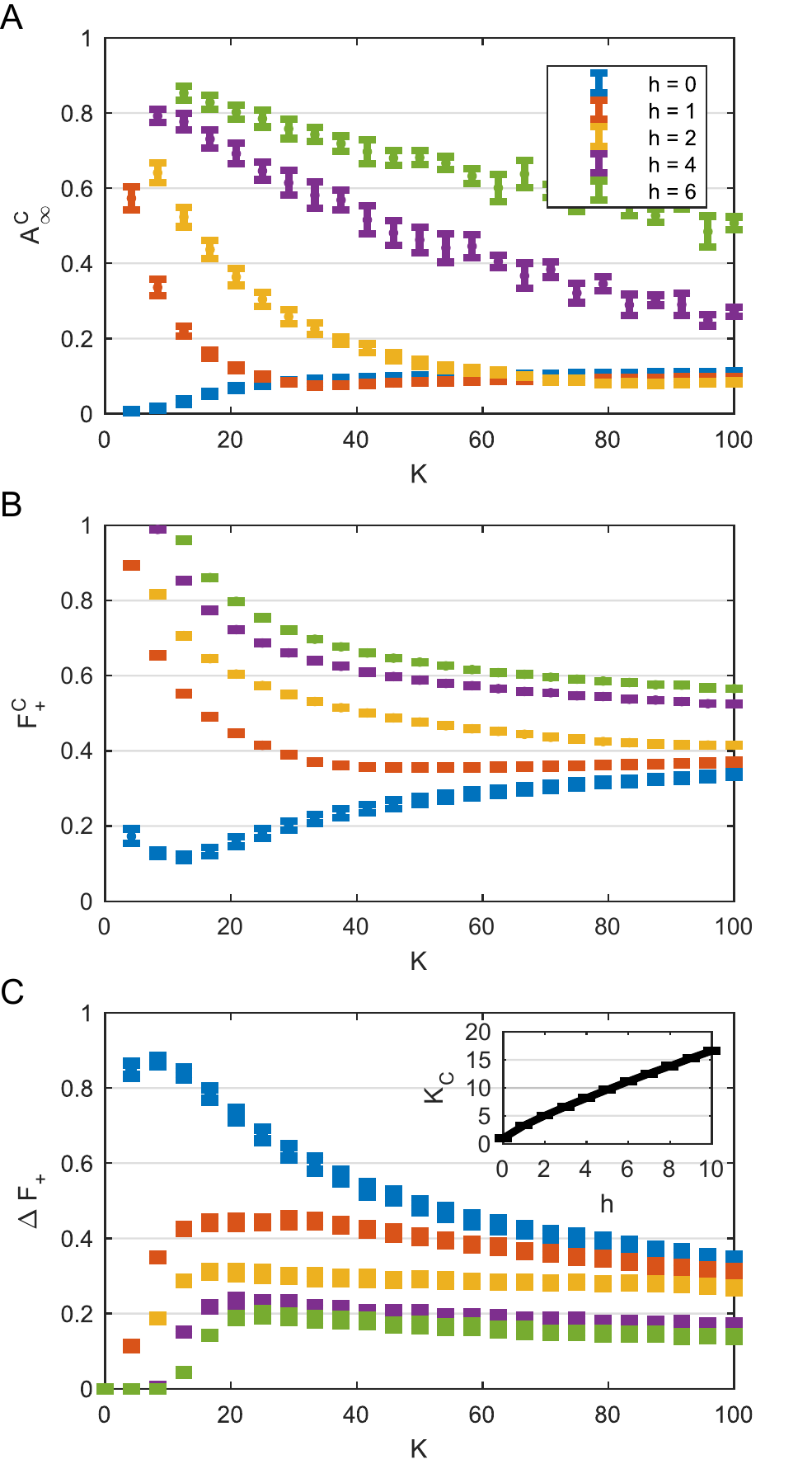}
\par\end{centering}

\centering{}\caption{\label{fig:DynamicRange}\textbf{Dynamic range as a function of the
average degree $K$ for $0\leq h\leq6$ A.} Minimum stable activity
$A_{\infty}^{C}$ that an RTN can sustain. \textbf{B. }Critical $F_{+}^{C}$
needed for the network to sustain activity. \textbf{C.} Non-frozen
range $\Delta F_{+}$ where the network produces non-frozen dynamics
$0<A_{\infty}<1$. Inset: Minimum degree $K_{C}$ needed for sustained
activity. Parameters are $n=10^{3}$ and $A\left(0\right)=0.95$,
and each point is the average of $10^{2}$ simulations. The color
code is the same for all figures.}
\end{figure}

Different combinations of topological and dynamical parameters can
yield different values of $A_{\infty}$. With fixed $h$ and $K$
there is a critical $F_{+}^{C}=F_{+}^{C}\left(K\right)$ required
for $A_{\infty}>0$, and the corresponding minimum activity $A_{\infty}^{C}=A_{\infty}\left(F_{+}^{C}\right)$.
There is also a critical degree $K^{C}=K^{C}\left(h\right)$, which
is the minimum connectivity required for $A_{\infty}>0$ with $F_{+}=1$.
All of these quantities place boundaries on the activity of RTNs.
If $K<K_{C}$, or $F<F_{+}^{C}$, no activity can be maintained by
the network. More importantly, $A_{\infty}^{C}$ is the minimum activity
the network can maintain, and the knowledge of it is necessary for
using Eq.\ \ref{eq:MF-final-inverse} to generate RTNs with desired
activity. In Fig.\ \ref{fig:DynamicRange} we show $A_{\infty}^{C}$
, $F_{+}^{C}$ and $K_{C}$ (inset) for $0\leq h<6$. While $A_{\infty}^{C}$
is low for low $h$, it can be quite high for high $h$. We obtain
$A_{\infty}^{C}=0.51$ for $K=100$ and $h=6$, meaning more than
half of the network must be active in order for it to maintain activity
with any $F_{+}$.

It is interesting to explore the range of values $A_{\infty}$ can
take by varying one parameter, as complete control of all parameters
may not be feasible in certain situations. Lets us define the non-frozen
range $\Delta F_{+}$ of $A_{\infty}$ as $\Delta F_{+}=F_{2}(K,h)-F_{1}(K,h)$,
where $F_{2}$ is the highest and $F_{1}$ the lowest $F_{+}$that
produces non-frozen dynamics ($0<A_{\infty}<1$) for a certain set
of $(K,h)$. If there is no $0<A_{\infty}<1$ possible we set $\Delta F_{+}=0$.
In Fig.\ \ref{fig:DynamicRange}C we show $\Delta F_{+}$.\textbf{
}As we increase both $K$ and $h$ the range $\Delta F_{+}$ decreases,
meaning the non-frozen dynamics gets compressed into a smaller range
of values of $F_{+}$. The dynamic range $\Delta A_{\infty}$ (i.e.,
the range of $A_{\infty}$ the network can have by varying $F_{+}$)
is given by $\Delta A_{\infty}=1-A_{\infty}^{C}$. As $K$ increases
$A_{\infty}^{C}$ decreases, with the exception of $h=0$ and low
$K$. This means that in order to maintain a high dynamic range it
is advantageous for the network to be highly connected.

\subsection{Adaptive networks with threshold dynamics}

It is interesting to study how the RTN dynamics deals with an evolving
topology. As we will see in Sec. III, the dynamics is fairly independent
on the precise network topology (Fig.\ \ref{fig:MF_A}C), making it
a good candidate for a system that must be stable under changes to the
topology. In the brain, during development a massive pruning process
removes a large portion of the neuronal synapses \cite{Huttenlocher1987}.
Inspired by this, we study how a simple link-removal adaptation rule
can shape the dynamics of the network. 

The algorithm consists of removing an excitatory in-link of node $i$
if its average activity $\bar{A_{i}}$ is $\bar{A_{i}}\geq\alpha$,
and removing an inhibitory in-link otherwise. While it can be expected
that the algorithm can drive the individual node activity towards
$\alpha$, it is not obvious whether it preserves the RTN dynamics
or drives it towards a chaotic state. The algorithm, and its implications,
are described in more detail in Appendix B.

\begin{figure}
\begin{centering}
\includegraphics[width=8.6cm]{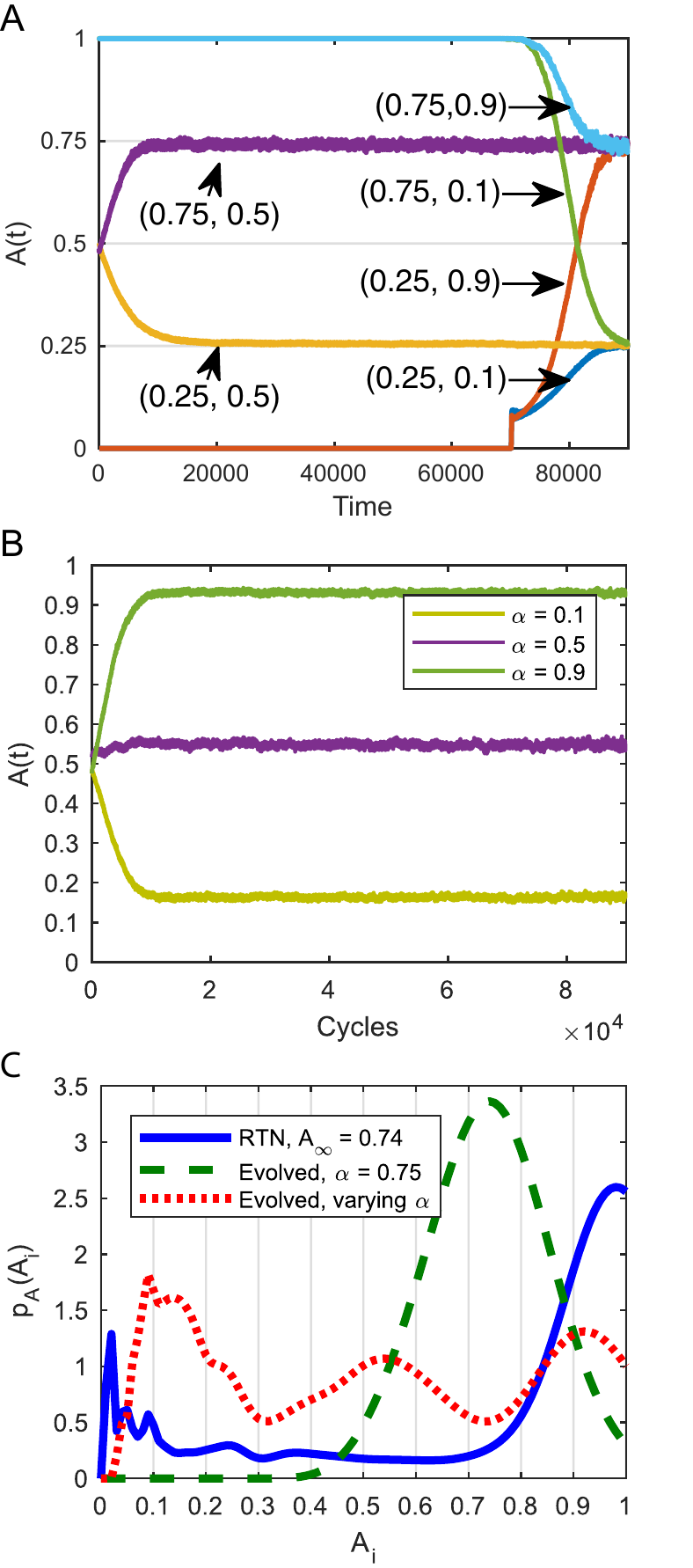}
\par\end{centering}

\centering{}\caption{\textbf{\label{fig:AdaptiveNet}Network activity with an adaptive
algorithm. A.} Evolution with a single $\alpha$ for the network.
The arrows indicate the parameters $\left(\alpha,F_{+}^{0}\right)$.
\textbf{B.} Evolution with multiple $\alpha\in\{0.1,0.5,0.9\}$ within
the same network. \textbf{C.} Activity distribution between nodes
$p_{A}(\bar{A_{i}})$ for the RTN (full line) with $F_{+}=0.54$ and
$A_{\infty}=0.74$, evolved network with $\alpha=0.75$ and $F_{+}^{0}=0.5$
(dashed line) and the evolved network from (B) (dotted line). Other
parameters are $n=10^{3}$ , $K=10^{2}$ and $h=0$.}
\end{figure}

Our simple algorithm consistently drives the global, stable activity
to $A_{\infty}\rightarrow\alpha$. (Fig.\ \ref{fig:AdaptiveNet}A)
for very different $\alpha$ and initial fraction $F_{+}^{0}$ of
excitatory links Thus, it can be used to evolve a network with thresholding
dynamics towards a certain activity level with only local information.
A natural extension is to allow different parameters $\alpha_{i}$
for each node. In Fig.\ \ref{fig:AdaptiveNet}B we divide the $n=10^{3}$
nodes into $3$ groups with $\alpha\in\{0.1,0.5,0.9\}$, and show
that the algorithm can lead to the co-existence of groups with very
different activity rates within the same network.

The adaptive algorithm allows us to control not only $A_{\infty}$,
but the node activity distribution $p_{A}(\bar{A_{i}})$. This is
particularly useful because the RTN dynamics with a random (ER) network
results in a bimodal $p_{A}(\bar{A_{i}})$. In other words, a high
number of nodes is either always on ($A_{i}=1$) or always off ($A_{i}=0$),
which can be undesirable. By evolving the network, we can obtain more
a Gaussian activity distribution. This is shown in Fig.\ \ref{fig:AdaptiveNet}C,
where and RTN with $F_{+}=0.54$ and an evolved network with $\alpha=0.75$
have the same $A_{\infty}=0.74$, but very different $p_{A}(\bar{A_{i}})$.
The $p_{A}(\bar{A_{i}})$ of the evolved network from Fig.\ \ref{fig:AdaptiveNet}B
displays three peaks centered around each of the $\alpha\in\{0.1,0.5,0.9\}$.
Thus, for large enough networks, this can in principle be used to
obtain an arbitrary activity distribution $p_{A}(\bar{A_{i}})$ within the network.

\section{Mean-field Theory}

\subsection{Activity $A_{\infty}$\label{sub:Mean-field-approximation}}

The annealed approximation was introduced by Derrida \& Pomeau \cite{Derrida1986b}
and it is a useful tool to study the dynamics of RTNs. The idea behind
it is to ignore temporal correlations between nodes, making each node
independent. This is equivalent to shuffling the network links at
each timestep. When we take into account the effect of $F_{+}$, the
activity at time $t+1$ is given by
\begin{equation}
A(t+1)=\sum_{k=1}^{n-1}p_{k}\sum_{m=h+1}^{k}\binom{k}{m}A(t)^{m}\left(1-A(t)\right)^{k-m}P_{+}(m)\label{eq:MF-full}
\end{equation}
where $P_{+}(m)$ is given by 
\begin{equation}
P_{+}(m)=\sum_{l=\lfloor\frac{m+h}{2}\rfloor+1}^{m}\binom{m}{l}F_{+}^{l}\left(1-F_{+}\right)^{m-l}\label{eq:MF-full-Pplus}
\end{equation}
with $p_{K}$ denoting the degree distribution of the network, and
$\lfloor x\rfloor$ the floor function of $x$. In principle, 
Eq.\ \ref{eq:MF-full} can be solved graphically to yield $A_{\infty}$.
It compares favorably with numerical simulations for the dynamics
of RTNs without a self-regulating state \cite{Aldana2011}. An interesting
feature of Eq.\ \ref{eq:MF-full} is that it predicts more than two
fixed points $A_{\infty}$ for some cases. For instance, for $K=25$,
$h=2$ and $F_{+}=0.6$ with an ER topology the fixed points are $A_{\infty}\in\{0,0.10,0.49\}$.
Nonetheless, we were unable to find more than one stable non-zero
fixed point ($A_{\infty}=0.49$ in this case) in the explored parameter
space. 

\begin{figure}
\begin{centering}
\includegraphics[width=8.6cm]{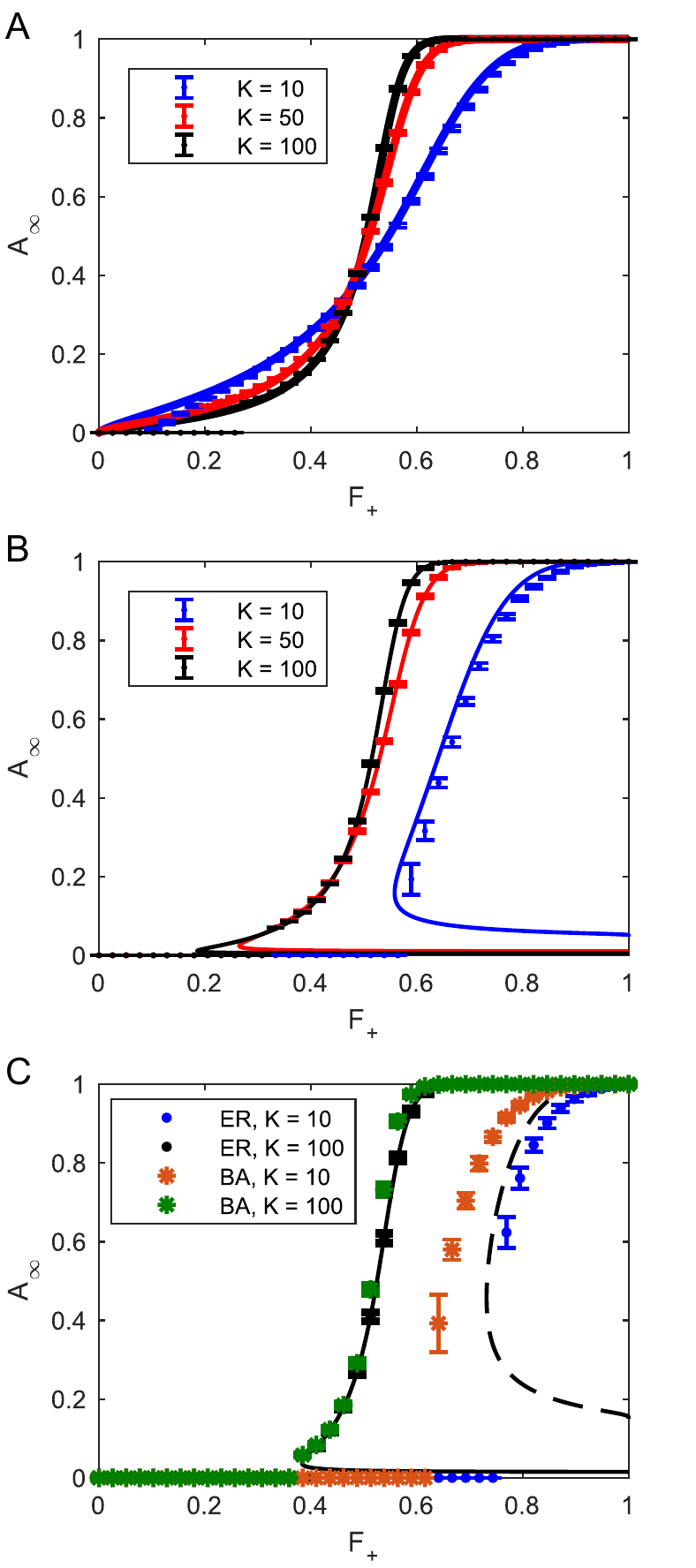}
\par\end{centering}

\centering{}\caption{\label{fig:MF_A}\textbf{Comparison between simulations (points) and
the mean-field approximation (lines) for $A_{\infty}$. A.} Results
for ER networks with $h=0$ and $K\in\{10,50,100\}$. \textbf{B.}
Same as (A) for $h=1$. \textbf{C.} Comparison for ER and BA networks
with the approximation for $K=10$ (solid line) and $K=100$ (dashed
line). Parameters are $n=10^{3}$ and $A\left(0\right)=0.95$, and
each point is the average of $10^{2}$ simulations.}
\end{figure}

The annealed approximation has some quirks, however. It demands the
knowledge of the full degree distribution, which is not relevant in
the high-degree regime (Fig.\ \ref{fig:MF_A}C). Care must also be
taken when numerically evaluating it, as naively computing $\binom{m}{l}F_{+}^{l}\left(1-F_{+}\right)^{m-l}$can
lead to machine precision errors. More importantly, it does not easily
tell us how to generate a network with a determined value of $A_{\infty}$.
In this section we obtain a simplified version of Eq.\ \ref{eq:MF-full}
with the aim to facilitate its application to the study of RTNs.

Let us first look into Eq.\ \ref{eq:MF-full-Pplus}, as $P_{+}(m)$
is the most important term of Eq.\ \ref{eq:MF-full}. We can use the
identity
\begin{equation}
I_{p}(n+1,N-n)=\sum_{i=n+1}^{N}\binom{N}{n}p^{i}(1-p)^{N-i}
\end{equation}
to write $P_{+}(m)$ in terms of the regularized incomplete beta function
$I_{z}(a,b)$, which is defined for non-integer $a$ and $b$. By
approximating$\lfloor x/2\rfloor\thickapprox x/2-1/4$, we can write

\begin{equation}
P_{+}(m)=I_{F_{+}}\left(\frac{m+h+3/2}{2},\frac{m-h+1/2}{2}\right)\label{eq:MF-Pplus}
\end{equation}
As already mentioned, in the high-degree regime the influence of the
degree distribution is small. We can then substitute the sum over
$p_{K}$ in Eq.\ \ref{eq:MF-full} for the average degree $K$. If
we assume that $P_{+}(m)$ is a slow-varying function, we can remove
it from the innermost sum in Eq.\ \ref{eq:MF-full} and substitute
$m$ by its average value $\bar{m}=KA_{\infty}$. We can then write

\begin{equation}
A_{\infty}=\sum_{m=h+1}^{K}\left\{ \binom{K}{m}A_{\infty}^{m}\left(1-A_{\infty}\right)^{K-m}\right\} P_{+}\left(KA_{\infty}\right)
\end{equation}
For high degree $K$, we can approximate $\sum_{m=h+1}^{K}\binom{K}{m}A_{\infty}^{m}\left(1-A_{\infty}\right)^{K-m}\approx1,$
and using Eq. \ref{eq:MF-Pplus} we write
\begin{equation}
A_{\infty}=I_{F_{+}}\left(\frac{KA_{\infty}+h+3/2}{2},\frac{KA_{\infty}-h+1/2}{2}\right)\label{eq:MF-final}
\end{equation}
 The usefulness of Eq.\ \ref{eq:MF-final} lies in the fact that $I_{z}\left(a,b\right)$
is a standard function in most algebra packages. Therefore, Eq.\ \ref{eq:MF-final}
can be quickly solved graphically or numerically with minimal effort.
More importantly, the inverse of $I_{X}(a,b)=Y$, $I_{Y}^{-1}(a,b)=X$
is also a standard function. We can use it to write
\begin{equation}
F_{+}=I_{A_{\infty}}^{-1}\left(\frac{KA_{\infty}+h+3/2}{2},\frac{KA_{\infty}-h+1/2}{2}\right)\label{eq:MF-final-inverse}
\end{equation}
which gives us the necessary value for $F_{+}$ to generate a RTN
with degree $K$ and activity $A_{\infty}$. This allows us to control
the activity $A_{\infty}$ of the network by manipulating the balance
between excitatory and inhibitory links.

In Fig.\ \ref{fig:MF_A} we compare the results from Eq.\ \ref{eq:MF-final-inverse}
(plotted with inverted axis) and numerical simulations. Our simplified
approximation matches well the simulation results in most cases. However,
it fails to capture the transition between $A_{\infty}=0$ and $A_{\infty}>0$,
producing a second low $A_{\infty}$ fixed point that is not observed
in Eq.\ \ref{eq:MF-full}. This is caused by our assumption that $P_{+}(m)$
is slow-varying breaking down at the transition between $A_{\infty}=0$
and $A_{\infty}>0$. Our simplification of $p_{k}=\delta_{k,K}$ can
also produce divergences for low $K$ and high $h$, where the degree
distribution of the network can significantly change $A_{\infty}$. 
This is evident in Figure \ref{fig:MF_A}C, where we compare $A_{\infty}$for
networks with Erdős–Rényi (ER) and Barabási-Albert
(BA) topologies \cite{Erdos1959,Barabasi1999} and $h=2$. As $K$
increases, however, both converge to the same $A_{\infty}$. For $h>2$,
both ER and BA networks converge to the same $A_{\infty}$ even with
$K=10$ (not shown).

It is important to note that for high $K$ the approximation fails
in one direction: predicting $A_{\infty}>0$ for a network with $A_{\infty}=0$.
When the network has a non-zero $A_{\infty}$ the results of our approximation
are quite accurate to describe the fixed point activity of the network.
Therefore, if we know that the network can maintain activity and is
well-connected, we can use Eq.\ \ref{eq:MF-final} and \ref{eq:MF-final-inverse}
to predict its properties. In other words, our approximation is applicable
as long as $A_{\infty}>A_{\infty}^{C}$, which is shown in Fig.\ \ref{fig:DynamicRange}A.

\subsection{Sensitivity $\lambda$}

\begin{figure}
\begin{centering}
\includegraphics[width=8.6cm]{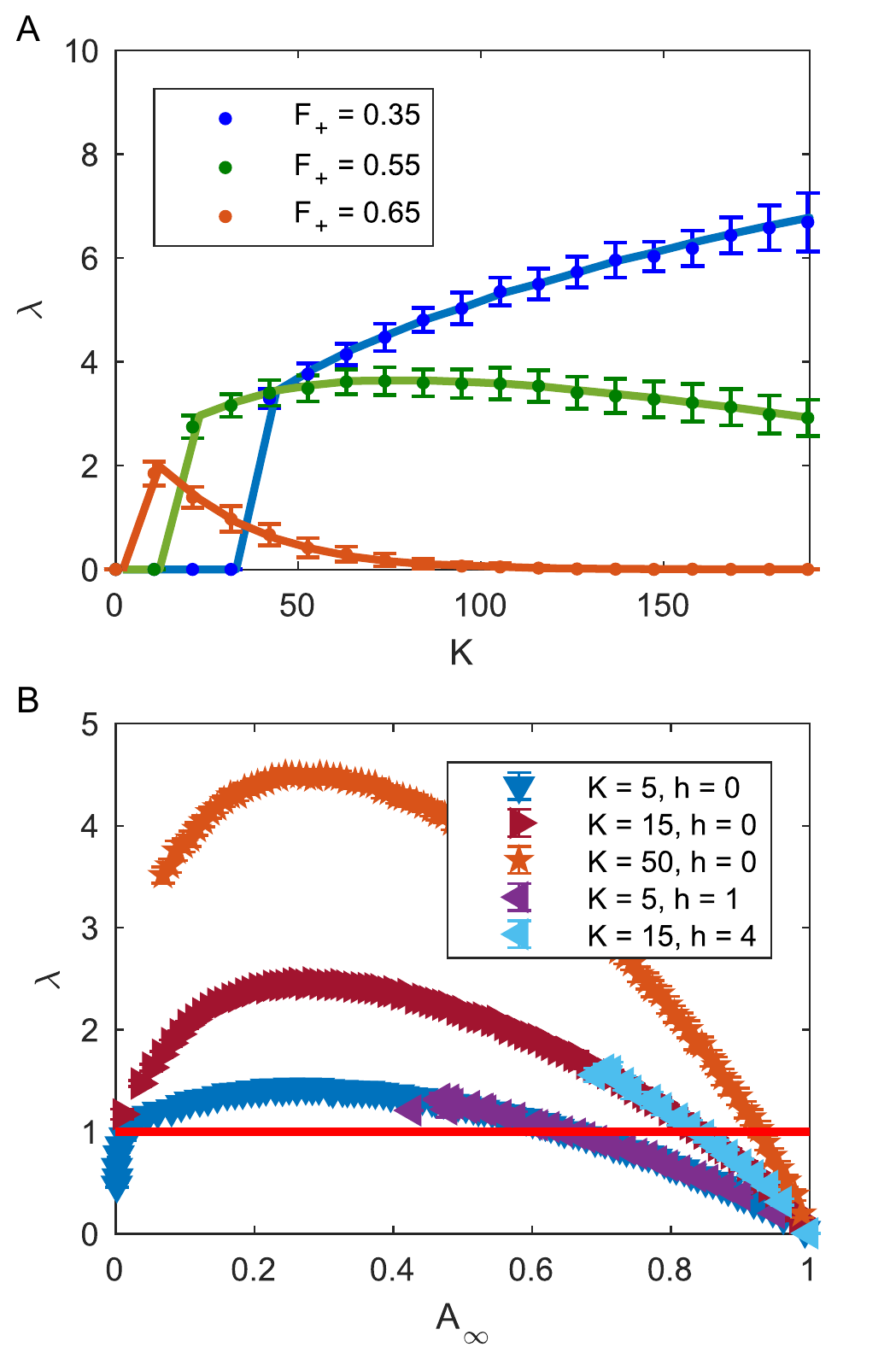}
\par\end{centering}

\centering{}\caption{\label{fig:sensitivity}\textbf{Sensitivity $\lambda$ of RTNs. A.
}Comparison between simulations (points) and the mean-field approximation
(lines) for $h=1$. \textbf{B.} Relationship between $A_{\infty}$
and $\lambda$ for varying $K$ and $h$. The red line corresponds
to critical dynamics ($\lambda=1$). Parameters are $n=10^{3}$ and
$A\left(0\right)=0.9$, and each point is the average of $10^{2}$
simulations.}
\end{figure}

The sensitivity of RTNs was studied in \cite{Aldana2011}. The model used,
however, has long-term correlations resulting in a complicated mean-field
approximation. In this section we propose a simpler mean-field approximation
of $\lambda$ for the dynamics defined in Eq.\ \ref{eq:def-rtn}, and
study its relationship to the network activity $A_{\infty}$.

The basic idea of the approximation is to look into the contribution
of a node $j$ in the input sum $S_{i}(t)=\sum_{k}w_{ik}\sigma_{k}(t)$
of node $i$. Let us define the input sum of $i$ at time $t$ without
considering $j$ as $S_{i}'=\sum_{k\neq j}w_{ij}\sigma_{k}(t)$. For
a threshold of $h$, a flip of $\sigma_{j}(t)$ can only cause a flip
of $\sigma_{i}(t+1)$ if $S_{i}'(t)=h$ or $S_{i}'(t)=h+1$. If $P(m)$
is the probability of node $i$ having $m$ active neighbors between
$k-1$ in-neighbors, the probability $P_{k}(j,i)$ of node $j$ inducing
a flip on node $i$ is given by
\begin{eqnarray}
P_{k}(j,i) & = & \sum_{m=h}^{k-1}P(m)\biggl\{ P(S_{i}'=h)P_{k}(j,i|\,S_{i}'=h)+\nonumber \\
 &  & P(S_{i}'=h+1)P_{k}(j,i|\,S_{i}'=h+1)\biggr\}\label{eq:p_flip1_lambda}
\end{eqnarray}
where 
\begin{equation}
P(m)=\binom{k-1}{m}A_{\infty}^{m}\left(1-A_{\infty}\right)^{k-1-m}\label{eq:pm_lambda}
\end{equation}
For $m$ active nodes $S_{i}'=h$ is only possible if $m+h$ is even
and $m\geq h$. In this case we have
\begin{equation}
P(S_{i}'=h)=\binom{m}{(m+m)/2}F_{+}^{(m+h)/2}\left(1-F_{+}\right)^{(m-h)/2}
\end{equation}
Likewise, $S_{i}'=h+1$ is only possible if $m+h$ is odd and $m\geq h+1$.
In this case
\begin{equation}
P(S_{i}'=h+1)=\binom{m}{\frac{m+h+1}{2}}F_{+}^{(m+h+1)/2}\left(1-F_{+}\right)^{(m-h-1)/2}\label{eq:ps_hp1}
\end{equation}
If $S_{i}'=h$, node $j$ can only alter the state of $i$ if $w_{ij}=+1$.
If $S_{i}'=h+1$ however, node $j$ can flip $i$ only if $w_{ij}=-1$.
In other words, $P_{k}(j,i|\,S_{i}'=h)=F_{+}$ and $P_{k}(j,i|\,S_{i}'=h+1)=1-F_{+}$.
Substituting Eq. \ref{eq:pm_lambda}-\ref{eq:ps_hp1} into Eq. \ref{eq:p_flip1_lambda},
we have
\begin{eqnarray}
P_{k}(j,i) & = & \sum_{m=h}^{k-1}\binom{k-1}{m}A^{m}(1-A)^{K-1-m}\times\\
 &  & F_{+}^{(m+h)/2}\left(1-F_{+}\right)^{(m-h)/2}\gamma_{h}\left(m,F_{+}\right)\nonumber 
\end{eqnarray}
where
\begin{equation}
\gamma_{h}\left(m,F_{+}\right)=\begin{cases}
\binom{m}{(m+h)/2}F_{+} & \mbox{, \ensuremath{m+h} even}\\
\binom{m}{(m+h+1)/2}\sqrt{F_{+}\left(1-F_{+}\right)} & \mbox{, \ensuremath{m+h} odd}
\end{cases}\label{eq:gamma-MF-S}
\end{equation}
The sensitivity $\lambda$ is the mean value of $P_{k}(j,i)$:
\begin{equation}
\lambda=\sum_{k=1}^{n-1}p_{k}^{in}P_{k}(j,i)\label{eq:S-full}
\end{equation}
 where $p_{k}^{in}$ denotes the in-degree distribution of the network.
If we approximate $p_{k}^{in}\rightarrow K$, we have our final result
\begin{eqnarray}
\lambda & = & K\sum_{m=h}^{K-1}\binom{K-1}{m}A_{\infty}^{m}(1-A_{\infty})^{K-1-m}\label{eq:sensitivity-final}\\
 &  & \times F_{+}^{(m+h)/2}\left(1-F_{+}\right)^{(m-h)/2}\gamma_{h}\left(m,F_{+}\right)\nonumber 
\end{eqnarray}
where $\gamma_{h}\left(m,F_{+}\right)$ is given by Eq.\ \ref{eq:gamma-MF-S}.
In Fig.\ \ref{fig:sensitivity}A we compare Eq.\ \ref{eq:sensitivity-final}
to simulation results for $h=1$. Our approximation provides a good
match to the results.

As both $\lambda$ and $A_{\infty}$ depend on the same set of parameters,
they cannot be freely chosen. We can then use $\lambda$ as a constraint
on $A_{\infty}$. This can be done by numerically solving the coupled
system made of Eq.\ \ref{eq:MF-full} and Eq.\ \ref{eq:S-full}. However,
in the high $K$ regime we can substitute Eq.\ \ref{eq:MF-final-inverse}
into Eq.\ \ref{eq:sensitivity-final} to obtain $\lambda=\lambda\left(A_{\infty},K,h\right)$.
In Fig.\ \ref{fig:sensitivity}B we show the relationship between $\lambda$
and $A_{\infty}$ for varying $K$ and $h$. The network is insensitive
to perturbation for $A_{\infty}=0$ and $1$. Between
these two extremes $\lambda$ is a concave function of $A_{\infty},$
with higher $K$ resulting in higher $\lambda$. While $h$ creates
regions with $\lambda=0$ (Fig.\ \ref{fig:DynamicRange}B), it does
not significantly change the value of $\lambda>0$. There exists a
critical point ($\lambda=1$) for most $K$ and $h$ and high $A_{\infty}$.
However, if $K$ is low (and $h$ is also low, to allow $A_{\infty}>0$)
another critical point will appear with low $A_{\infty}.$

\section{Discussion}
At the core of our study is the idea that Random Threshold Networks are robust to changes
in the topology, and that their activity can be controlled by balancing
excitation and inhibition. 
Here, we found a much richer dynamics for varying fractions of excitatory and inhibitory links than for the special case of equal balance ($F_+ = 0.50$).
The stable network activity covers a large and non-trivial range of values, and we find that network sensitivity and activity depend on the same factors, and cannot be freely chosen. 

More importantly we find that, for well-connected networks, criticality requires an excess of excitatory connections, and is only possible in a narrow band of inhibition. Our model results pose interesting questions w.r.t. inhibition in the brain. GABAergic neurons \textit{in vitro} are known to shift from having an excitatory to inhibitory effect during development \cite{Ben-Ari2002}. In our model, the shift results in subcritical dynamics early on being a necessary step to reach criticality, which is compatible with results of avalanches of activity \textit{in vitro} \cite{Pasquale2008,Levina2017}. Evidence points towards no GABA shift \textit{in vivo}, however, with GABAergic neurons always being inhibitory \cite{Kirmse2015, Valeeva2016}. In this situation, our model predicts that a constant level of inhibition is required for criticality during network growth. This is compatible with the finding that the fraction of GABAergic neurons is constant during development \textit{in vivo} \cite{Sahara2012}. Thus, our results on the influence of inhibition on the critical point possibly provide independent support for the hypothesis of a near-critical dynamics in cortical tissue.

Using a simple adaptive algorithm, we increased our control from the global average to the activity distribution within the network. The model can also be
extended in other ways in order to generate specific topologies. For
instance, a link creation rule can be used to balance the link pruning,
or link shuffling to maintain the degree distribution. The topological
evolution rules can also be changed in order to control other properties
of the network. Scale-free networks with controllable activity can in principle be generated both through growth \cite{Barabasi1999} and pruning \cite{Schneider2011}.

Overall our RTN model, using simple threshold units, is able to generate rich dynamics with a phase transition. It does not depend on details of the network topology, and has both inhibitory interactions and stable, ceaseless dynamics. This can make the RTN an interesting minimal model to explore mechanisms of neural dynamics, as the brain has inhibition, an evolving topology and a dynamics with input integration and thresholding.

\begin{acknowledgments}
JPN thanks the financial support of the São Paulo Research Foundation (FAPESP) under grants 2012/18550-4 and 2013/15683-6, and of the Brazilian National Council for Scientific and Technological Development (CNPq) under grant 206891/2014-8.
MAM thanks the financial support of FAPESP under grants 2016/06054-3 and 2016/01343-7, and CNPq under grant 302049/2015-0. 
JAB thanks the financial support of FAPESP under grant 2016/04783-8.

\end{acknowledgments}

\bibliographystyle{unsrt}
\bibliography{References}

\newpage{}

\appendix

\section{Dependence on $A(0)$}

\begin{figure}
\begin{centering}
\includegraphics[width=8.6cm]{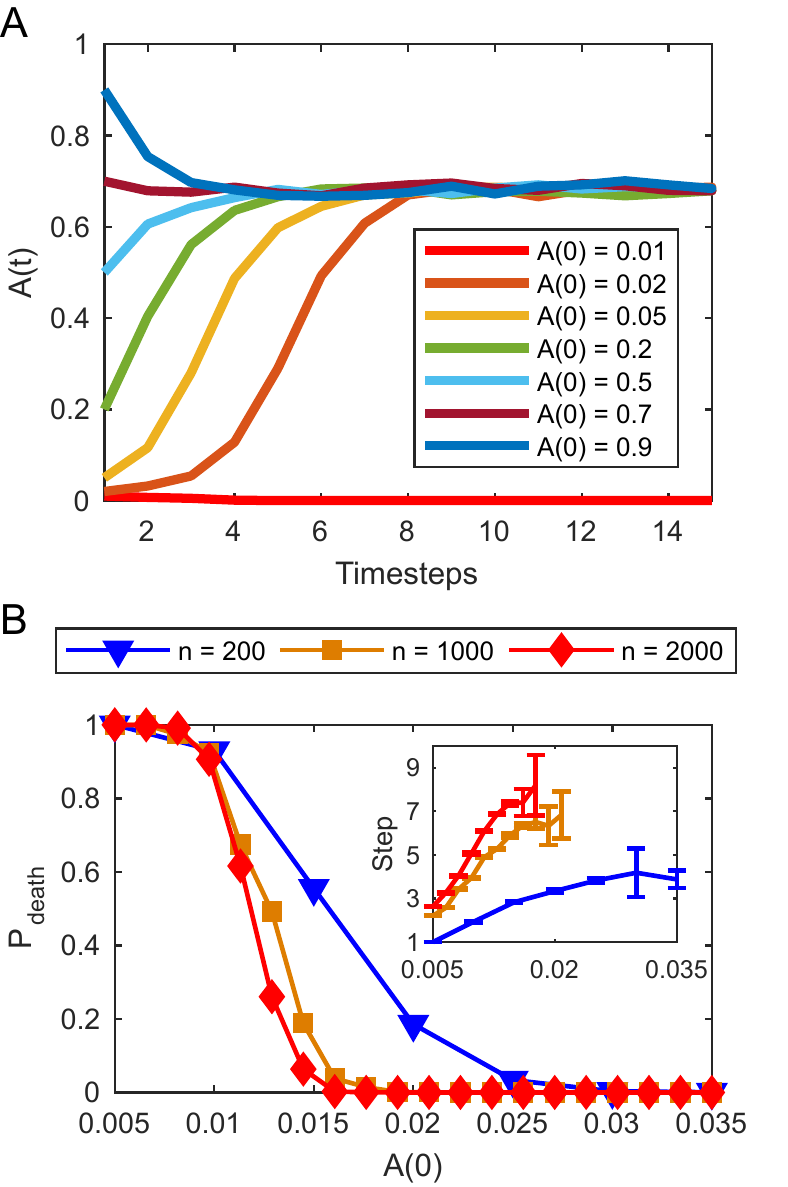}
\par\end{centering}

\begin{centering}
\caption{\textbf{\label{fig:EffectA0}Effect of $A(0)$ on the dynamics. A.}
Effect of different $A(0)$ on a network with $n=1000$. \textbf{B.}
Probability $P_{\text{death}}$ of the dynamics dying out for $n\in\{200,1000,2000\}$.
Inset: Average step of death for the simulations that ended in $A=0$.
Each point is an average from $20$ networks and $10^{3}$ simulations
each. Other parameters are $K=25$, $F_{+}=0.6$ and $h=1$ for both
cases.}

\par\end{centering}

\end{figure}

Which fixed point ($A_{\infty}=0$ or $A_{\infty}>0$ if it exists)
the dynamics stays at depends on the input $A(0)$ used to activate
the network. A critical input $A_{0}^{C}$ is needed for ceaseless
dynamics, below which the network quickly goes to $A_{\infty}=0$.
This is exemplified in Fig.\ \ref{fig:EffectA0}A, where the dynamics
dies out for $A(0)=0.01$ but not for $A(0)\geq0.02$. The value of
$A_{0}^{C}$ can be obtained numerically from a mean-field approximation
(Sec. IIIA), corresponding to the $n\rightarrow\infty$ case. For
any realization of the dynamics, however, there is a probability of
it dying out even if $A(0)>A_{0}^{C}$. This probability depends chiefly
on the value of $A(t)$, so if $A_{\infty}$ is high the dynamics
is only likely to die at the beginning and with a low $A(0)$. The
time it takes for the dynamics to reach $A_{\infty}$ depends predominantly
on $A(0)$, but also on factors such as the degree distribution $p_{K}$
of the network and $\lambda$. In Fig.\ \ref{fig:EffectA0}B we show
the dependency of the probability $P_{\text{death}}$ of extinction
(i.e. $A(t)=0$) on $A(0)$ as a function of $n$. As $n\rightarrow\infty$,
$P_{\text{death}}$ resembles a step function. In order to sidestep
this issue, in the main text we always activate the network with a
high $A(0)=0.90$ or $A(0)=0.95$.

\section{Adaptive algorithm}

Here we describe in more detail the adaptive algorithm used in Sec.
IID. The average activity of node $i$ is defined as $\bar{A_{i}}=\sum_{t=0}^{t=\tau}\sigma_{i}(t)/\tau$,
where $\tau=10$ defines a moving average. In the adaptive process,
each $\bar{A_{i}}$ is compared to a parameter $\alpha_{i}$ and change
its connectivity according to a certain set of rules. The important
question is then which set of rules, combined with a specific distribution
of $\bar{A_{i}}$, can produce a certain topology and network activity.
In other words, the problem amounts to the exploration of the class
of adaptive threshold networks with local, activity-based topological
evolution. Here we follow the simple algorithm:
\begin{enumerate}
\item If $A(t)=0$ activate a fraction $A_{0}$ of the network nodes.
\item Run the dynamics of Eq.\ \ref{eq:def-rtn} for $\tau$ timesteps, and
choose a random node $i$.
\item If $\bar{A_{i}}>\alpha_{i}$, remove a random positive in-link of
$i.$ Otherwise remove a random negative in-link. If there is no suitable
link for removal, choose another $i$. 
\item Iterate from step 1. 
\end{enumerate}
The above algorithm can only be run a finite number of cycles. Since
a single link is removed after each cycle, we can set the average
degree $K$ of the evolved network by running the algorithm $\tau_{cycle}=n\left(K_{0}-K\right)$
cycles, where $K_{0}$ is the average degree of the initial network.
The choice of $\tau$ is not important, since $A(t)$ stabilizes quickly.
This leaves us with only two evolutionary parameters, $\alpha$ and
$\tau_{cycle}=n\left(K_{0}-K\right)$. If we start from a fully connected
network, the in-degree distribution $p_{k}^{in}$ of the evolved network
is given by

\begin{equation}
p_{k}^{in}=\binom{n(n-1-K)}{n-1-k}\left(\frac{1}{n}\right)^{n-1-k}\left(1-\frac{1}{n}\right)^{n(n-1-K)-(n-1-k)}
\end{equation}
and the out-degree distribution $p_{k}^{out}$ is given by 
\begin{equation}
p_{k}^{out}=\binom{n}{k}\left(\frac{K}{n-1}\right)^{k}\left(1-\frac{K}{n-1}\right)^{n-k}
\end{equation}
While both distributions are binomial, the in-degree distribution
is wider than the out-degree.
\end{document}